\documentclass[aps,prb,superscriptaddress,twocolumn,longbibliography]{revtex4-1}
\usepackage{bbm}
\usepackage{graphicx}
\usepackage{dcolumn}
\usepackage{bm}
\usepackage{subfigure}
\usepackage{amsmath}
\usepackage{feynmf}
\usepackage{hyperref}
\usepackage{color}

\usepackage{attachfile}

\newcommand{\bk}{\boldsymbol k}

\newcommand{\zb}{\color {black}}
\usepackage{times}

\begin{document}
\title{Field-sensitive dislocation bound states in two-dimensional $d$-wave altermagnets}

\author{Di Zhu}
\affiliation{Guangdong Provincial Key Laboratory of Magnetoelectric Physics and Devices,
State Key Laboratory of Optoelectronic Materials and Technologies, School of Physics, Sun Yat-Sen University, Guangzhou 510275, China}

\author{Dongling Liu}
\affiliation{Guangdong Provincial Key Laboratory of Magnetoelectric Physics and Devices, State Key Laboratory of Optoelectronic Materials and Technologies, School of Physics, Sun Yat-Sen University, Guangzhou 510275, China}

\author{Zheng-Yang Zhuang}
\affiliation{Guangdong Provincial Key Laboratory of Magnetoelectric Physics and Devices, State Key Laboratory of Optoelectronic Materials and Technologies, School of Physics, Sun Yat-Sen University, Guangzhou 510275, China}

\author{Zhigang Wu}
\affiliation{Quantum Science Center of Guangdong-Hong Kong-Macao Greater Bay Area (Guangdong), Shenzhen 508045, China}

\author{Zhongbo Yan}
\email{yanzhb5@mail.sysu.edu.cn}
\affiliation{Guangdong Provincial Key Laboratory of Magnetoelectric Physics and Devices, State Key Laboratory of Optoelectronic Materials and Technologies, School of Physics, Sun Yat-Sen University, Guangzhou 510275, China}

\date{\today}

\begin{abstract}
When a two-dimensional $d$-wave altermagnet is grown on a substrate,
the  interplay of momentum-dependent spin splittings arising from
altermagnetism and Rashba spin-orbit coupling gives rise to a nodal band structure with band degeneracies
enforced by a $C_{4z}\mathcal{T}$ symmetry. If the $C_{4z}\mathcal{T}$ symmetry is broken by an exchange field,
we find that the band degeneracies are immediately  lifted, leading to a
topological band structure characterized by nontrivial strong and weak topological indices. Remarkably,
both the strong topological index  and the $Z_{2}$-valued weak topological indices depend sensitively on the direction of the exchange field. As a consequence of the bulk-defect correspondence, we find that the unique dependence of
weak topological indices on the exchange field in this system dictates that the presence or absence of
topological bound states at lattice dislocations also depends sensitively on the direction of the exchange field.
When the substrate is an $s$-wave superconductor, we find that a similar
dependence of band topology on the exchange field gives rise to field-sensitive
dislocation Majorana zero modes. As topological
dislocation bound states are easily detectable by scanning tunneling microscopy,
our findings unveil a promising experimental diagnosis of altermagnetic materials among an ever growing list of candidates.
\end{abstract}

\maketitle

\section{Introduction}

Altermagnetism (AM) has attracted increasing interest recently as a collinear magnetic order
with salient properties distinct from the conventional collinear ferromagnetism and antiferromagnetism\cite{Libor2022AMa,Libor2022AMb,Libor2020AM,Hayami2019AM,Hayami2020AM,Yuan2020AM,
Yuan2021AM,Mazin2021,Shao2021NC,Ma2021AM,Liu2022AM,Feng2022AM,Betancourt2023,Mazin2023AM,Turek2022AM,
Hariki2023AM,Bhowal2024AM,Zhou2023AM,Han2024AM,Chen2023AM,Xiao2023AM,Jiang2023AM}.
In real space, the magnetic moments of an altermagnet are collinearly arranged
to form a N\'{e}el order, just like an antiferromagnet. However, unlike
the antiferromagnets the two sublattices with opposite magnetic moments in a magnetic
unit cell of the altermagnet cannot be mapped to each other
by the combined symmetry operations of time reversal and inversion/translation.
Instead, they are mapped to each other by the combined symmetry operations of time reversal
and rotation/mirror\cite{Libor2022AMa,Libor2022AMb}. A remarkable consequence of this difference in symmetry is that the
AM leads to momentum-dependent spin-splitting electronic band structures but maintains
symmetry-compensated zero net magnetization. Thus, it is not only distinct from  ferromagnetism
which leads to momentum-independent spin splitting and finite magnetization, but also distinguishes itself from antiferromagnetism which results in
degenerate band structures.
Notably, the spin splitting induced by AM can reach
the order of $1$eV, and the symmetry pattern of the spin splitting is rich,
exemplified by the classification of AM into groups dubbed $d$-, $g$- and $i$-wave AM.
Excitingly, several materials, including insulating MnTe\cite{Osumi2024MnTe,Lee2024MnTe,Krempasky2024,Hajlaoui2024AM} and metallic CrSb\cite{Reimers2024,Ding2024CrSb,Yang2024CrSb,Zeng2024CrSb,Li2024CrSb},
have been experimentally confirmed to be altermagnets by high-resolution
angle-resolved photoemission spectroscopy (ARPES).
On the theoretical front,  first-principle \cite{Guo2023AM,Gao2023AM,Qu2024AM,Joachim2024AM,Bernardini2024AM,Sheoran2024AM,Marko2024AM,Zeng2024AM,Liu2024AM}
and model \cite{Leeb2024AM,Purnendu2023AM,Toshihiro2023AM,Roig2024AM,Pedro2024AM,Yu2024AM,Bose2024AM} calculations have predicted an ever growing list of candidates for AM. Furthermore, many studies have also shown that AM
can give rise to numerous interesting effects and phases,
such as giant and tunneling magnetoresistance\cite{Libor2022AMc},
diverse tunneling phenomena in superconductor/altermagnet
junctions\cite{Ouassou2023AM,Sun2023AM,Papaj2023AM,Beenakker2023AM,Cheng2024AM,Nagae2024,Giil2024AM,Das2024AM,Lu2024AM},
finite-momentum Cooper pairing\cite{Zhang2024AM,Sumita2023FFLO,Chakraborty2023AM},
unconventional superconductivity\cite{Zhu2023TSC,Wei2024AM,Brekke2023AM,Kristian2024AM,Chourasia2024AM},
various types of Hall effects\cite{Parshukov2024AM,Fang2023NHE,Attias2024AM,Hoyer2024AM}, and anisotropic RKKY interaction between magnetic impurities\cite{Lin2023AM,Amundsen2024RKKY}.

Because altermagnetic materials are often grown on substrates which can induce spin-orbit coupling (SOC),
theoretical studies of altermagnetic materials need to also take into account this effect. SOC is known as another basic mechanism giving rise to
momentum-dependent spin splitting and the interplay of AM and SOC
can result in band structures of rich topological properties\cite{Fernandes2024AM}. Thus far,
the $d$-wave AM has been the focus of most of the theoretical studies. It has been shown
that the band structure of an intrinsic $d$-wave altermagnet normally has spin-polarized
Dirac points in two dimensions, and the further presence of SOC can gap out the Dirac points and
result in first-order topological insulator phases\cite{Antonenko2024AM,Parshukov2024AM}.
Furthermore, when intrinsic superconductivity occurs in a two-dimensional (2D) $d$-wave altermagnet with Rashba SOC,
 both first-order and second-order topological superconductivity
are found to emerge\cite{Zhu2023TSC}. Lastly,
it has been shown that second-order topological insulators or superconductors
can be obtained in hybrid systems composed
of $d$-wave altermagnets and first-order topological insulators or superconductors\cite{Ghorashi2023AM,Li2023AMHOTSC,Li2024AMHOTI,Li2024AMTSC,Ezawa2024AM}.
In all these studies, the nontrivial momentum-space topology of the bulk bands is manifested in
the presence of topological boundary states, dictated by the bulk-boundary correspondence.
Interestingly,  real-space topological defects, which are ubiquitous in materials and are characterized by real-space
topological invariants, can also reflect the momentum-space
topology in a distinct way, known as the bulk-defect correspondence\cite{Teo2010,Chiu2015RMP,Yao2017floquet}.

The types of topological defects in materials are diverse\cite{Mermin1979,Kleman2008dislocation}.
For lattice topological defects, disclinations and dislocations are two classes that attract
particular interest in the context of topological phases\cite{Teo2017defect,Lin2023defect,Asahi2012defect,Buhler2014,Hughes2014dislocation,Slager2014dislocation,
Fernando2014dislocation,Guido2018dislocation,Roy2021dislocation,Rodrigo2020dislocation,Schindler2022,Hu2024dislocation}.
In the seminal work of Ran, Zhang and Vishwanath\cite{ran2009one}, it was discovered that a lattice dislocation can harbor a pair of 1D gapless helical modes in 3D
topological insulators; the topological criterion for the existence of these
topological dislocation modes is given by $\textbf{B}\cdot\textbf{M}_{\nu}=\pi \,(\text{mod} \,2\pi)$,
where $\textbf{B}$ refers to the Burgers vector which is the real-space topological
invariant characterizing the dislocation,  and $\textbf{M}_{\nu}=\sum_{i=1}^{3}\nu_{i}\textbf{G}_{i}/2$.
Here $\nu_i=\{0,1\}$ is the weak topological indices defined on the high symmetry planes
at the Brillouin zone's boundary\cite{fu2007a}
and $\textbf{G}_{i}$ is the reciprocal lattice vectors. $\textbf{M}_{\nu}$ acts like a time-reversal invariant momentum.
Later,
it was demonstrated that the topological criterion can also be generalized
to two dimensions  and is applicable to topological superconductors as well\cite{Ran2010dislocation}.
In two dimensions,  when the topological criterion is fulfilled,
the topological dislocation modes are 0D bound states.
As an application, it was shown that the presence or absence of dislocation bound states in a 2D topological insulator
with $C_{4z}$ rotation symmetry
can serve as a bulk probe to diagnose the location of band inversion\cite{Juricic2012dislocation}.

\begin{figure}[t]
\centering
\includegraphics[width=0.45\textwidth]{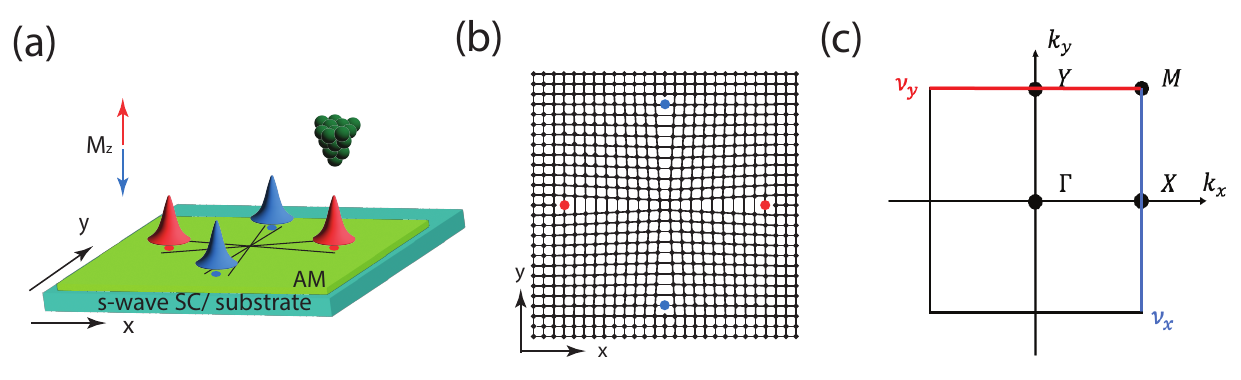}
\caption{(a) A schematic diagram of
field-sensitive dislocation bound states and their experimental measurement by scanning tunneling microscopy.
The 2D altermagnetic material grown on an insulator substrate or an $s$-wave superconductor has two pairs of
dislocations with perpendicular Burgers vectors.
The color correspondence between the bell-shaped surfaces denoting topological dislocation bound states and the arrows denoting the exchange field's direction illustrates the fact that which pair of the dislocations carries topological bound states relies on the exchange field's direction.
(b) Details of the dislocation configuration. The two blue (red) dots correspond to the cores of a pair of dislocations with Burgers vector $\textbf{B}=\pm \bm{e_x}$ ($\pm \bm{e_y}$).
The two pairs of dislocations with perpendicular Burgers vectors form a cross-shaped defect. (c)
2D Brillouin zone. $\boldsymbol{\Gamma}=(0,0)$, $\textbf{X}=(\pi,0)$, $\textbf{Y}=(0,\pi)$ and $\textbf{M}=(\pi,\pi)$ are
four high symmetry points. Two weak topological indices, $\nu_{x}$ and $\nu_{y}$, which determine the presence or absence of
topological dislocation bound states,  are defined on the blue and red lines
at the Brillouin zone's boundary.
}\label{fig1}
\end{figure}

In this paper, we investigate topological dislocation modes in a 2D $d$-wave altermagnet grown on a substrate.
Because of the structure asymmetry, Rashba SOC becomes an important factor
in the determination of the band structure in the altermagnet. Although the $d$-wave AM breaks the time-reversal symmetry ($\mathcal{T}$)
and $C_{4z}$ rotation symmetry, it respects their combination, the $C_{4z}\mathcal{T}$ symmetry. As
the Rashba SOC also respects this symmetry, the cooperation of $d$-wave AM and Rashba SOC leads to
a unique spin-splitting band structure with spin textures and Berry curvatures
respecting this symmetry too\cite{Zhu2023TSC}. Furthermore, despite the existence
of spin splitting at generic momenta in the Brillouin zone, the band structure
has nodal points at the two $C_{4z}\mathcal{T}$ invariant momenta.
The $C_{4z}\mathcal{T}$-symmetric band structure serves as the indication of a critical phase,
since a weak perturbation breaking the  $C_{4z}\mathcal{T}$ symmetry
can gap out the nodal points and lead to topological gapped phases. As we are interested in these gapped phases,
we consider the presence of an additional exchange field which will break the $C_{4z}\mathcal{T}$ symmetry.
Remarkably, we find that both the strong topological index (Chern number)
and the $Z_{2}$-valued weak topological indices characterizing the gapped band structure
depend sensitively on the exchange field's direction. As a result,
whether a dislocation harbors topological bound states  hinges on the exchange field's
direction.  Furthermore, when the substrate is an $s$-wave superconductor,
we find that a similar dependence of band topology on the exchange
field gives rise to field-sensitive dislocation Majorana zero modes.

The rest of the paper is organized as follows. In Sec.\ref{II}, we describe the
theoretical model and analyze the dependence of topological indices on
the exchange field. By considering a cross-shaped defect consisting
of two pairs of dislocations with perpendicular Burgers vectors,
we illustrate that the presence of topological bound states at the dislocation cores
depends sensitively on the exchange field's direction.
In Sec.\ref{III}, we generalize this analysis and show
that field-sensitive dislocation Majorana zero modes can be obtained
when the substrate is an $s$-wave superconductor.
In Sec.\ref{IV}, we discuss our findings and conclude the paper.

\section{Dislocation bound states in 2D metal with $d$-wave AM and Rashba SOC}\label{II}

We first consider a 2D altermagnet grown on an insulator substrate and subject to a perpendicular exchange/Zeeman field,
as illustrated in Fig.\ref{fig1}(a). In this paper,  we do not distinguish between Zeeman field and exchange field in terms of terminology. In the absence of defects, the tight-binding Hamiltonian is given by\cite{Zhu2023TSC} $H=\sum_{\bm{k}}c_{\bm{k}}^{\dag}\mathcal{H}_{0}(\bm{k})c_{\bm{k}}$
with $c_{\bm{k}}^{\dag}=(c_{\bm{k},\uparrow}^{\dag},c_{\bm{k},\downarrow}^{\dag})$ and
\begin{eqnarray}
\mathcal{H}_{0}(\bm{k})&=&-2t(\cos{k_x}+\cos{k_y})\sigma_0+2\lambda(\sin{k_y}\sigma_x-\sin{k_x}\sigma_y)\nonumber\\
&&+[2t_{\rm AM}(\cos{k_x}-\cos{k_y})+M_z]\sigma_z,\label{Hamiltonian}
\end{eqnarray}
Here $\sigma_{i}$ are Pauli matrices acting on the spin degrees of freedom. {\zb This two-band Hamiltonian can emerge, for example, in 
the Lieb lattice. To be specific, when the sublattices at the centers of the $x$- and $y$-directional edges have opposite magnetic moments, the electrons from
the nonmagnetic atoms at the vertices of the Lieb lattice will experience an effective $d$-wave altermagnetic exchange field\cite{Brekke2023AM}. If one considers only the effective Hamiltonian for the nonmagnetic atoms,  the first hopping term (the $t$ term)
and the $d$-wave AM term (the $t_{\rm AM}$ term) in $\mathcal{H}_0$ will be obtained. The second term in $\mathcal{H}_0$ denotes the Rashba SOC originating from structure asymmetry, and the last $M_{z}$ term
accounts for a momentum-independent exchange field induced by  an external perpendicular magnetic field or a ferromagnetic insulator
substrate whose magnetization is perpendicular to the plane.} 
Throughout the lattice constant is set to unity for notational simplicity, and without loss of generality the parameters
$t$, $\lambda$ and $t_{\rm AM}$ are assumed to be non-negative for the convenience of discussion.

\begin{figure}[t]
\centering
\includegraphics[width=0.45\textwidth]{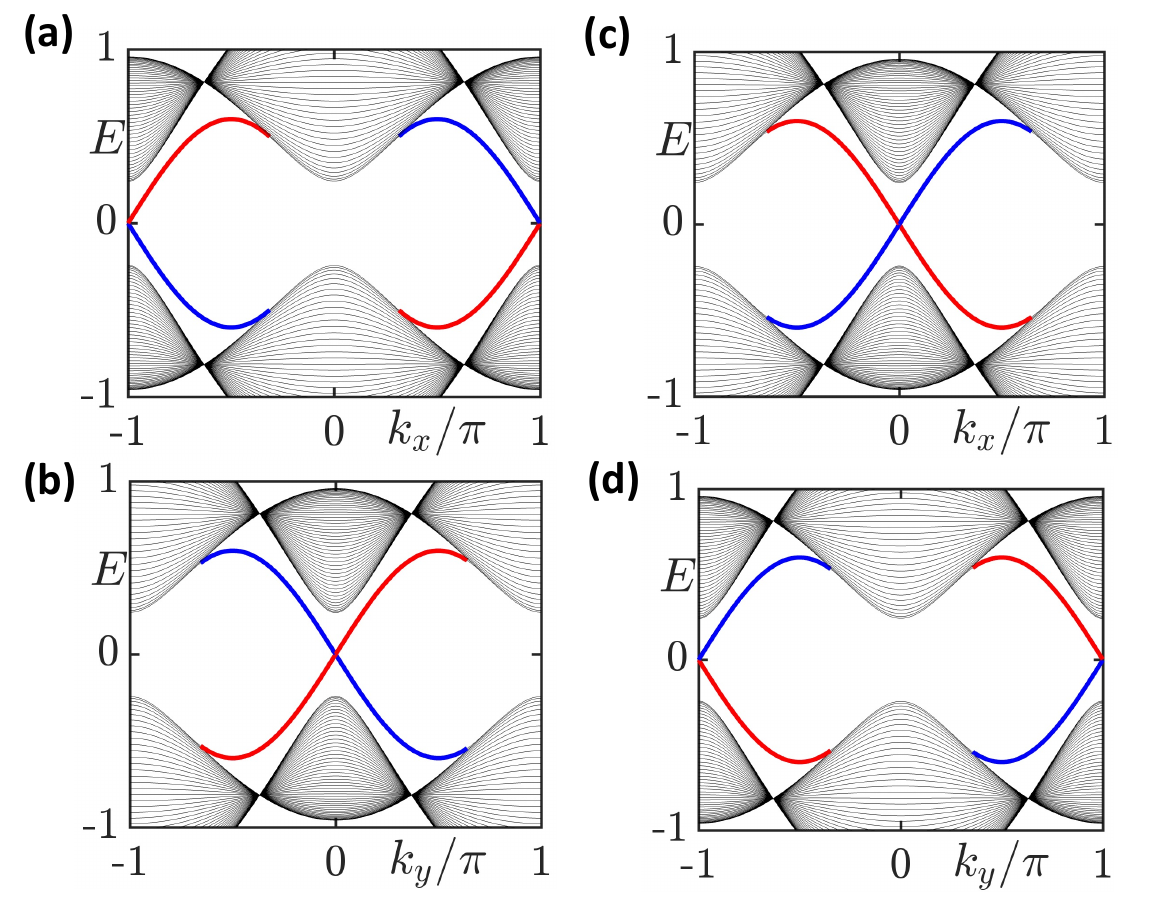}
\caption{Energy spectrum for a sample of ribbon geometry. For the top row, the system
takes periodic (open) boundary conditions in the $x$ ($y$) direction.
The bottom row is just the opposite. In (a) and (b), $M_z=0.24$, $C_{-}=-1$. The chiral
edge state propagates clockwise on the boundary, and the edge-state spectrum crossings
occur at $k_{x}=\pi$ and $k_{y}=0$, indicating $(\nu_{x},\nu_{y})=(1,0)$.
In (c) and (d), $M_z=-0.24$, $C_{-}=1$. The chiral
edge state becomes anticlockwise, and the edge-state spectrum crossings
are switched  to $k_{x}=0$ and $k_{y}=\pi$, indicating $(\nu_{x},\nu_{y})=(0,1)$.
The red and blue solid lines in
(a) and (c) [(b) and (d)] refer to the chiral edge states on the top and bottom $y$-normal (left and right $x$-normal) edges.
Common parameters are $t=0$, $t_{\rm AM}=0.3$ and $\lambda=0.3$.
}\label{fig2}
\end{figure}

When $M_{z}=0$, the Hamiltonian has the $C_{4z}\mathcal{T}$ symmetry, even though the
$C_{4z}$ rotation symmetry ($C_{4z}=e^{i\frac{\pi}{4}\sigma_z}$) and time-reversal symmetry
($\mathcal{T}=-i\sigma_y\mathcal{K}$ with $\mathcal{K}$ the complex conjugate operator)
are independently broken by the $d$-wave AM. Because of this combined symmetry, the
two energy bands have Kramers degeneracies at the two $C_{4z}\mathcal{T}$-invariant momenta,
i.e., $\bm{\Gamma}$ and $\textbf{M}$\cite{Zhu2023TSC}. Once  $M_{z}$ becomes finite,
the $C_{4z}\mathcal{T}$ symmetry is broken and the band degeneracies at $\bm{\Gamma}$ and
$\textbf{M}$ are lifted, resulting in a finite energy gap between the two bands. As the
system does not have time-reversal symmetry, the gapped band structure
is characterized by the first-class Chern number. The Chern numbers characterizing
the two bands are given by\cite{qi2006QWZ}
\begin{eqnarray}
C_\pm&=&\pm\frac{1}{2\pi}\int_{BZ}\frac{\bm{d}(\bm{k})\cdot [\partial_{k_x}\bm{d}(\bm{k})\times \partial_{k_y}\bm{d}(\bm{k})]}{2|\bm{d}(\bm{k})|^{3}}d^2k\nonumber\\
&=&\pm
\begin{cases}
{\rm sgn}(M_z), &0<|M_z|<4t_{\rm AM},\\
0,&|M_z|>4t_{\rm AM}.
\end{cases}
\end{eqnarray}
where the subscript $+/-$ refers to the upper/lower band,
and $\bm{d}(\bk)=(2\lambda \sin k_{y},-2\lambda \sin k_{x},2t_{\rm AM}(\cos{k_x}-\cos{k_y})+M_z)$, with the components
$d_{i}(\bk)$ corresponding to the coefficient functions in front of the Pauli matrix $\sigma_{i}$ in Eq.~(\ref{Hamiltonian}). The dependence of Chern number on $M_{z}$ suggests that
an arbitrarily weak perpendicular exchange field renders the band structure topologically nontrivial.
Furthermore, the Chern number has a sensitive dependence on the direction of the exchange field.
Thus, a reversal of the direction of the exchange field will change the sign of the Chern number. As topological phases have bulk-boundary
correspondence, it is natural to expect that the topological boundary states will also have an intriguing dependence on
the exchange field.

{\zb By considering a sample of ribbon geometry (periodic boundary conditions in $x$ direction
and open boundary conditions in $y$ direction, or vise versa)
and calculating the energy spectrum  of the Hamiltonian in Eq.(\ref{Hamiltonian})}, we find as expected that
the chiral edge state  reverses its chirality when the exchange
field reverses its direction. What is surprising, however, is
that the momentum at which
the edge-state spectrum crosses undergoes a jump.
When $M_{z}>0$, the edge-state spectrum crossing occurs at
$k_{x}=\pi$ on the $y$-normal edges [Fig.\ref{fig2}(a)] and at $k_{y}=0$ on the $x$-normal edges [Fig.\ref{fig2}(b)].
Reversing the direction of the exchange field, the crossing momentum has a jump of half the reciprocal lattice vector. That is, the edge-state
spectrum crossing is shifted to
$k_{x}=0$ on the $y$-normal edges [Fig.\ref{fig2}(c)] and to $k_{y}=\pi$ on the $x$-normal edges [Fig.\ref{fig2}(d)]. Interestingly,
the crossing momenta on the $x$-normal edges and $y$-normal edge states are just switched upon reversing 
the exchange field. This switching behavior
can be understood through the $C_{4z}\mathcal{T}$ symmetry. To be specific, for the $x$-normal edge states under positive $M_z$ as shown in Fig. \ref{fig2}(a), applying the $C_{4z}\mathcal{T}$ symmetry operation gives the $y$-normal edge states under negative $M_z$ as shown in Fig. \ref{fig2}(d), and vise versa. Similarly, the spectra in Fig. \ref{fig2}(b) and Fig. \ref{fig2}(c) are connected by the same symmetry operation. As the edge-state spectrum crossings at time-reversal invariant momenta are connected to weak topological indices defined on 1D lower submanifolds of the Brillouin zone, namely,
high symmetry lines of the Brillouin zone, their dramatic jump
suggests that the weak topological indices of this Hamiltonian also have a sensitive dependence on the
direction of the exchange field.

The weak topological indices of interest in this work  are defined on the two high symmetry
lines at the boundaries of the 2D Brillouin zone, as illustrated in Fig.\ref{fig1}(c).
The two weak topological indices, labeled as $\nu_{x}$ and $\nu_{y}$, are $Z_{2}$-valued and
their parities determine
whether there is an odd number of edge-state spectrum crossings at $k_{x}=\pi$ and $k_{y}=\pi$ of the
respective boundary Brillouin zone.
Although the $C_{4z}\mathcal{T}$ symmetry is broken by the exchange field, the whole system
still has the $C_{2z}$ rotation symmetry described by the symmetry operator $C_{2z}=i\sigma_{z}$.
Owing to the existence of this crystalline symmetry,
the weak topological indices can be defined in terms of
the eigenvalues of the $C_{2z}$ operator at $C_{2z}$-invariant momenta.
Following the same spirit as the Fu-Kane formula for $Z_{2}$ invariants of
topological insulators\cite{fu2007a}, the explicit formulas for
the two $Z_{2}$-valued weak topological indices are given by
\begin{eqnarray}
(-1)^{\nu_{x}}=-\xi(\textbf{X})\xi(\textbf{M}),(-1)^{\nu_{y}}=-\xi(\textbf{Y})\xi(\textbf{M}).
\end{eqnarray}
where $\xi(\bm{k}_{R})$ is the
$C_{2z}$ operator's eigenvalue for the lower-band eigenstate at the
$C_{2z}$-invariant momentum $\bm{k}_{R}\in\{\textbf{X},\textbf{Y},\textbf{M}\}$, i.e.,
$C_{2z}|u(\bm{k}_{R})\rangle=\xi(\bm{k}_{R})|u(\bm{k}_{R})\rangle$. Because these eigenvalues take values of $\pm i$,
a factor of ``$-1$''  is introduced on the right hand side of the above equations. A
straightforward calculation reveals
\begin{eqnarray}
(\nu_{x},\nu_{y})=\left\{\begin{array}{cc}
                    (1,0), & 0<M_z<4t_{\rm AM}, \\
                    (0,1), & -4t_{\rm AM}<M_z<0, \\
                    (0,0), & |M_z|>4t_{\rm AM}.
                  \end{array}\right.
\end{eqnarray}
The above result shows explicitly that the two weak topological indices
switch their values when the exchange field in the weak-field regime ($|M_{z}|<4t_{\rm AM}$)
reverses its direction. Based on the two weak topological indices, the momentum $\textbf{M}_{\nu}$
reflecting the band topology is given by $\textbf{M}_{\nu}=\pi(\nu_{x}\bm{e_{x}}+\nu_{y}\bm{e_{y}})$.

\begin{figure}[t]
\centering
\includegraphics[width=0.45\textwidth]{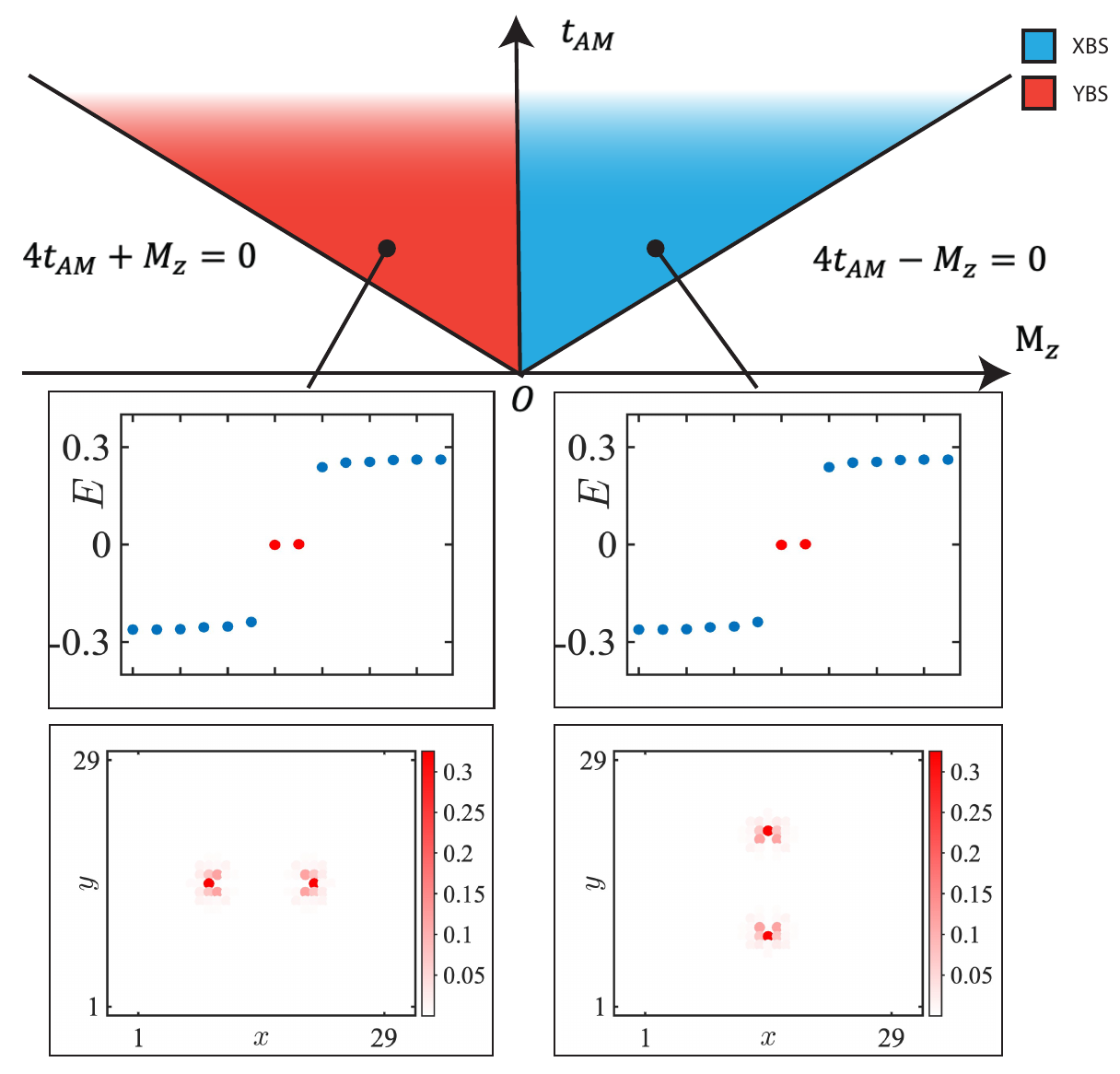}
\caption{\zb{Phase diagram determined via weak topological indices. The red and blue regions
correspond to $(\nu_{x},\nu_{y})=(0,1)$ and $(1,0)$, respectively. The two insets in the middle are energy spectra for a system with
lattice size $N_{x}\times N_{y}=29\times29$ and periodic boundary conditions in both $x$ and $y$ directions.
The two insets at the bottom show the corresponding distributions of the dislocation bound states in the lattice.
The results show that, in the region with $(\nu_{x},\nu_{y})=(0,1)$,
dislocations with Burgers vector $\textbf{B}=\pm\bm{e_y}$ harbor bound states, while the ones with $\textbf{B}=\pm\bm{e_x}$ do not,
and the situation is just the opposite for the region with $(\nu_{x},\nu_{y})=(1,0)$.
For the left (right) inset, $t_{\rm AM}=0.3$ and $M_z=-0.24$ $(0.24)$. The values of
other parameters are $t=0$ and $\lambda=0.3$.
}}\label{fig3}
\end{figure}

Let us now consider the presence of dislocations in the system.
As aforementioned, the topological criterion for the presence of topological bound states at dislocations in 2D
is also $\textbf{B}\cdot \textbf{M}_{\nu}=\pi$ (mod $2\pi$)\cite{Ran2010dislocation}.
Because of the $\textbf{M}_{\nu}$'s sensitive dependence on the exchange field's direction,
whether a dislocation with a fixed Burgers vector carries a topological bound state will thereby also sensitively depend on the
exchange field's direction.

To verify this expectation, we place
a cross-shaped defect consisting of two pairs of
dislocations in the system [see Fig.\ref{fig1}(b)] and numerically diagonalize
the Hamiltonian under periodic boundary conditions in both $x$ and $y$ directions. In the case of $0<M_z<4t_{\rm AM}$ for which $(\nu_{x},\nu_{y})=(1,0)$,
we find that the pair of dislocations with Burgers vector $\textbf{B}=\pm\bm{e_x}$ harbor topological bound states while those with Burgers vector $\textbf{B}=\pm\bm{e_y}$ do not, as shown in Fig.\ref{fig3}.
The picture is just reversed when $-4t_{\rm AM}<M_z<0$, demonstrating that the
presence or the absence of topological bound states at a dislocation can be tuned
by simply adjusting the exchange field's direction.

\section{Field-sensitive dislocation Majorana zero modes}\label{III}

Controllability is a desired property in the application of
Majorana zero modes in topological quantum computation\cite{Aasen2016}. Following the same line of argument as before, we will show in this section
that field-sensitive Majorana zero modes can also be achieved by taking advantage of the AM.
To be specific, we now consider the senario where a 2D $d$-wave altermagnetic metal is grown on a fully-gapped $s$-wave superconductor [see Fig.\ref{fig1}(a)] and is assumed to inherit the $s$-wave superconductivity from the bulk superconductor though
the proximity effect.  Within the Bogoliubov-de Gennes (BdG) framework, the effective Hamiltonian is given by
$H=\frac{1}{2}\sum_{\bm{k}}\Psi_{\bm{k}}^\dagger \mathcal{H}_{\rm BdG}(\bm{k}) \Psi_{\bm{k}}$, where
$\Psi^\dagger(\bm{k})=(c^\dagger_{\bm{k},\uparrow},c^\dagger_{\bm{k},\downarrow},c_{-\bm{k},\uparrow},c_{-\bm{k},\downarrow})$
and
\begin{eqnarray}
\mathcal{H}_{\rm BdG}(\bm{k})&=&[-2t(\cos{k_x}+\cos{k_y})-\mu]\tau_z\sigma_0\nonumber\\
  &&+[2t_{\rm AM}(\cos{k_x}-\cos{k_y})+M_z]\tau_z\sigma_z\nonumber\\
  &&+2\lambda(\sin{k_y}\tau_0\sigma_x-\sin{k_x}\tau_z\sigma_y)+\Delta_s\tau_y\sigma_y.\label{HamiltonianBdG}
\end{eqnarray}
Here $\mu$ is the chemical potential,  $\Delta_{s}$ is the proximity-induced $s$-wave pairing amplitude and  the new set
of Pauli matrices $\tau_{i}$ acts on the particle-hole degrees of freedom.

We again first focus on the band topology of the BdG Hamiltonian. Since the $s$-wave pairing
does not break the time-reversal symmetry and $C_{4z}$ rotation symmetry, the BdG Hamiltonian also
respects the $C_{4z}\mathcal{T}$ symmetry when $M_{z}=0$. In a previous work, we
have shown that the $C_{4z}\mathcal{T}$ symmetry forbids a 2D gapped superconductor to
have a nonzero Chern number\cite{Zhu2023TSC}. Despite the absence of strong topology characterized by the Chern number, weak topology
is compatible with this symmetry and can be nontrivial when the system's parameters fulfill certain conditions\cite{Ghorashi2023AM}.
To see this, we again make use of the eigenvalues of the $C_{2z}$ operator, which is now given by
$C_{2z}=-i\tau_{z}\sigma_{z}$ due to the inclusion of particle-hole degrees of freedom. More specifically,
we define
\begin{eqnarray}
(-1)^{\nu_{x}}&=&(-1)^{N}\prod_{n=1}^{N}\xi_{n}(\textbf{X})\xi_{n}(\textbf{M}),\nonumber\\
(-1)^{\nu_{y}}&=&(-1)^{N}\prod_{n=1}^{N}\xi_{n}(\textbf{Y})\xi_{n}(\textbf{M}),
\end{eqnarray}
where $N=2$ denotes the number of bands below $E=0$, $n$ is the band index with
$E_{n}<0$, and
$\xi_{n}(\bm{k}_{R})$ is the $C_{2z}$ operator's eigenvalue for the negative-energy eigenstate at $\bm{k}_{R}\in\{\textbf{X},\textbf{Y},\textbf{M}\}$, i.e.,
$C_{2z}|u_{n}(\bm{k}_{R})\rangle=\xi_{n}(\bm{k}_{R})|u_{n}(\bm{k}_{R})\rangle$.
A straightforward calculation yields the weak topological indices at $M_{z}=0$,
\begin{eqnarray}
(\nu_{x},\nu_{y})=\left\{\begin{array}{cc}
                    (1,1), & 4t_{\rm AM}>\sqrt{\mu^{2}+\Delta_{s}^{2}}, \\
                    (0,0), & 4t_{\rm AM}<\sqrt{\mu^{2}+\Delta_{s}^{2}}.
                  \end{array}\right.
\end{eqnarray}
This result indicates that, despite the impossibility of chiral Majorana modes
when $M_{z}=0$, Majorana zero modes can be created at dislocations
with Burgers vectors equal to either $\pm\bm{e_x}$ or $\pm\bm{e_y}$.
Before explicitly showing this, we first complete the analysis
of the bulk topology  when $M_{z}\neq0$.

\begin{figure}[t]
\centering
\includegraphics[width=0.45\textwidth]{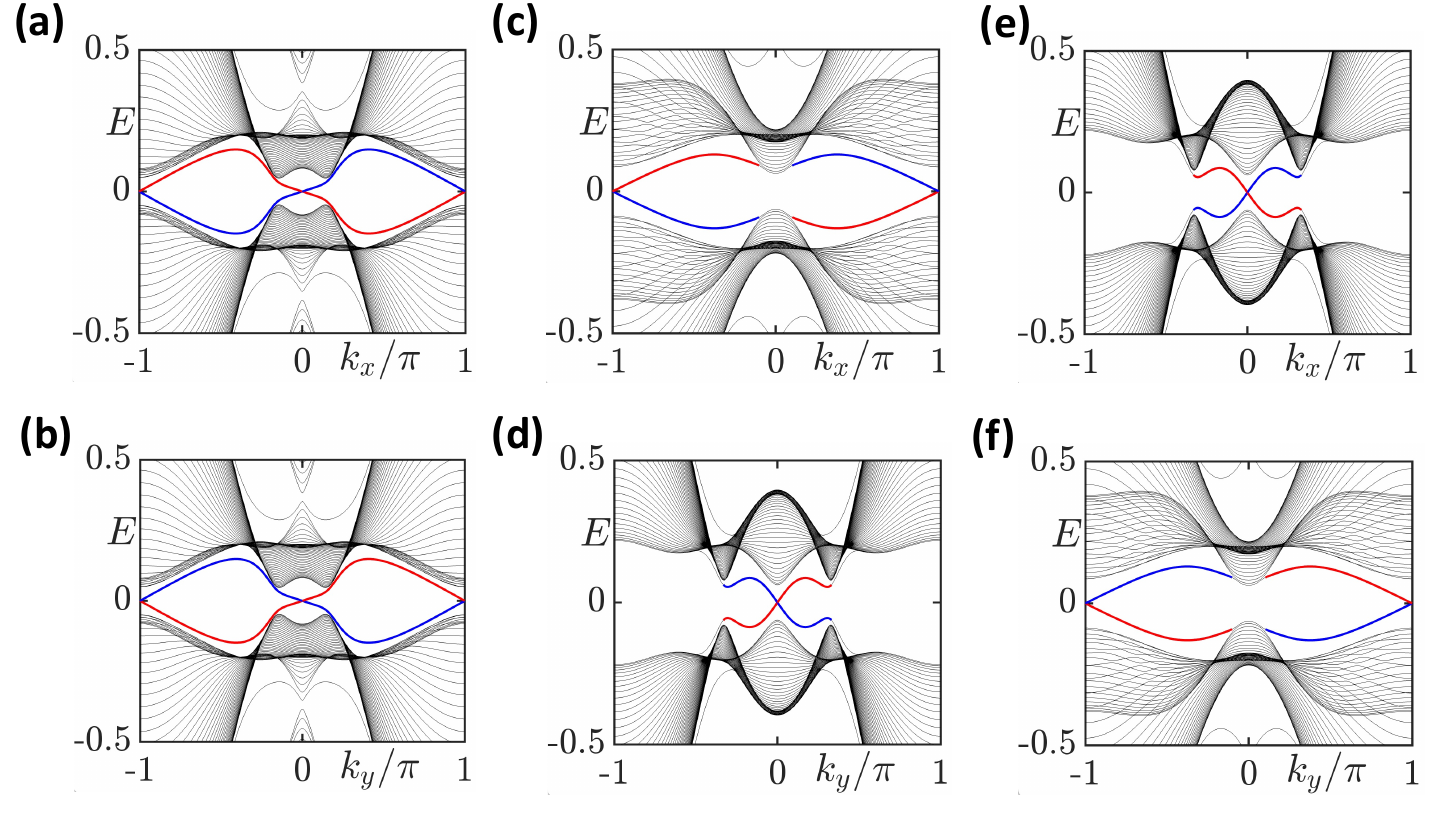}
\caption{BdG energy spectrum for a sample of ribbon geometry. The system takes open boundary directions
in the $y$ ($x$) direction and periodic boundary conditions in the $x$ ($y$) direction for the top (bottom) row.
In (a) and (b), $M_{z}=0$, $C=0$, and the locations of edge-state spectrum crossings suggest $(\nu_{x},\nu_{y})=(1,1)$.
In (c) and (d), $M_{z}=-0.3$, $C=-1$, and the locations of edge-state spectrum crossings suggest $(\nu_{x},\nu_{y})=(1,0)$.
In (e) and (f), $M_{z}=0.3$, $C=1$, and the locations of edge-state spectrum crossings suggest $(\nu_{x},\nu_{y})=(0,1)$.
Common parameters are $t=0.2$, $t_{\rm AM}=0.3$, $\lambda=0.3$, $\Delta_s=0.2$, and $\mu=-1.1$.
}\label{fig4}
\end{figure}

Similar to the previous nonsuperconducting case, a finite $M_{z}$ breaks
the $C_{4z}\mathcal{T}$ symmetry and
can induce topological superconducting phases characterized by nonzero Chern numbers.
The topological phase diagram can be easily determined since the
change of Chern number is associated with the close of bulk energy gap\cite{Sau2010TSC},
which takes place when $M_{z}=\{\pm\sqrt{(4t\pm \mu)^{2}+\Delta_{s}^{2}}$,
$\pm 4t_{\rm AM}\pm\sqrt{\mu^{2}+\Delta_{s}^{2}}$\}.
Here we are not interested in determining the complete phase diagram. Rather, we are interested in
 the region where the exchange field is weak, i.e., when $M_{z}$ is comparable to
$\Delta_{s}$, and when the chemical potential is close to the critical points
between the weak topological superconductor and trivial superconductors, i.e., when $\mu$ is close to
$\mu_{c,\pm}=\pm\sqrt{16t_{\rm AM}^{2}-\Delta_{s}^{2}}$. Without loss of generality,
we take $\mu=\mu_{c,-}+\delta$ where $\delta$ is a
small real positive constant. Then, increasing the exchange field's strength from zero,
the energy gap closes at
$\textbf{X}$ when $M_{z}=4t_{\rm AM}-\sqrt{\mu^{2}+\Delta_{s}^{2}}\simeq\delta$ (assuming $\Delta_{s}\ll t_{\rm AM}$),
and at $\textbf{Y}$ when $M_{z}=-4t_{\rm AM}+\sqrt{\mu^{2}+\Delta_{s}^{2}}\simeq-\delta$. The close of energy gap
at $\textbf{X}$ changes the total Chern number of the bands below $E=0$ from $C=0$ to $C=1$, and the weak topological indices $(\nu_{x},\nu_{y})$ from $(1,1)$ to
$(0,1)$. In contrast, the close of energy gap at $\textbf{Y}$ changes the total Chern number  of the bands below
$E=0$ from $C=0$ to $C=-1$, and
the weak topological indices $(\nu_{x},\nu_{y})$ from $(1,1)$ to
$(1,0)$.

In Fig.\ref{fig4}, we show the energy spectrum for a sample of ribbon geometry. In  Figs.\ref{fig4}(a) and \ref{fig4}(b),
we see that there are two counter-propagating gapless states on one edge and the edge-state spectrum crossing
occurs at time-reversal invariant momenta on both
$x$-normal and $y$-normal edges; this is consistent with $C=0$ and
$(\nu_{x},\nu_{y})=(1,1)$ for $M_{z}=0$ and $4t_{\rm AM}>\sqrt{\mu^{2}+\Delta_{s}^{2}}$.
In Figs.\ref{fig4}(c) and \ref{fig4}(d), the number of chiral edge states
and the locations of edge-state spectrum crossings suggest that $C=-1$ and
$(\nu_{x},\nu_{y})=(1,0)$, which is also consistent with the values as analyzed above.
In Figs.\ref{fig4}(e) and \ref{fig4}(f), the results show that a reversal of the exchange field's direction
reverses the chirality of the gapless edge state, and leads to a
switching of the locations of edge-state spectrum crossings on the $x$-normal and
$y$-normal edges. This again agrees with the sign change of the Chern number and the switching
of the values of $\nu_{x}$ and $\nu_{y}$ accompanying this process.

\begin{figure}[t]
\centering
\includegraphics[width=0.45\textwidth]{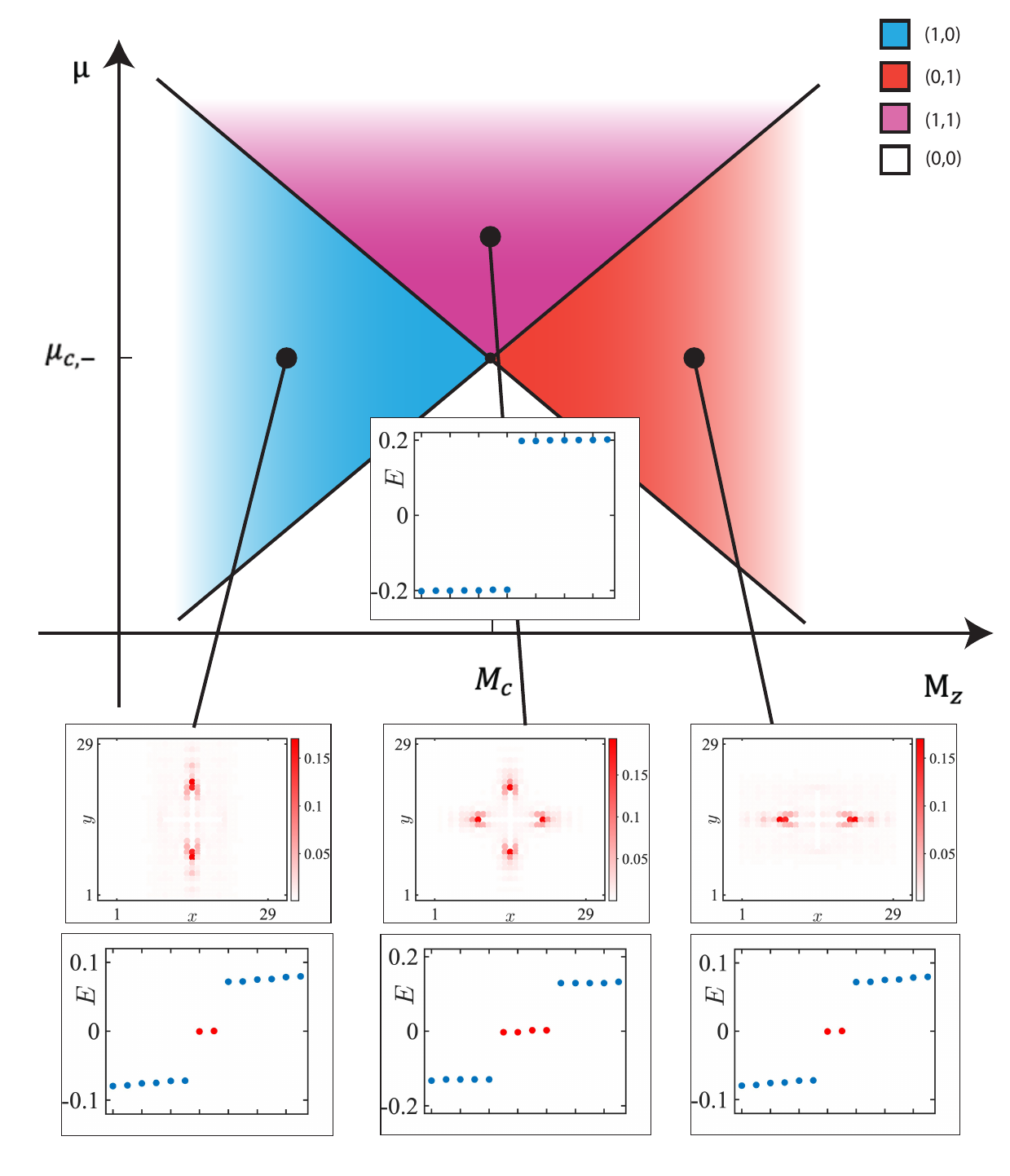}
\caption{\zb{Phase diagram around the critical point with $\mu_{c,-}=-\sqrt{16t_{\rm AM}^{2}-\Delta_{s}^{2}}$ and $M_c=0$.
The phase boundaries are given by $M_{z}=\pm(4t_{\rm AM}-\sqrt{\mu^{2}+\Delta_{s}^{2}})$. The weak
topological indices $(\nu_{x},\nu_{y})$ in the  blue, white, red and
purple regions take values of $(1,0)$, $(0,0)$, $(0,1)$ and (1,1), respectively. The
insets corresponding to the  four topologically distinct regions  show the number and distributions of Majorana zero modes
at the two pairs of dislocations with $\textbf{B}=\pm\bm{e_x}$ and $\pm\bm{e_y}$ in
a system with lattice size $N_{x}\times N_{y}=29\times29$ and periodic boundary conditions
in both $x$ and $y$ directions.
The parameters of the insets that correspond to the selected point in the left, right, top and bottom regions are $(\mu,M_z)=(-1.183,-0.3)$, $(\mu,M_z)=(-1.183,0.3)$, $(\mu,M_z)=(-0.6,0)$ and $(\mu,M_z)=(-1.4,0)$, respectively. Common parameters are $t=0.2$, $t_{\rm AM}=0.3$, $\lambda=0.3$ and $\Delta_s=0.2$.
}}\label{fig5}
\end{figure}

Placing a cross-shaped defect in the superconducting system and diagonalizing the
Hamiltonian under periodic boundary conditions in both directions, we find the presence of
Majorana zero modes at dislocations when the topological criterion $\textbf{B}\cdot \textbf{G}_{\nu}=\pi \,(\text{mod}\, 2\pi)$
is fulfilled. To be specific, when the system is a weak topological superconductor with
$(\nu_{x},\nu_{y})=(1,1)$ (indicated by the purple region of Fig.\ref{fig5}), we find that there are four Majorana zero modes
with their wave functions localized at the cores of the dislocations, as shown in the inset
on  top of  the purple region. The result suggests that dislocations with Burgers vector
$\textbf{B}=\pm\bm{e_x}$ and $\pm\bm{e_y}$
all harbor Majorana zero modes at their cores, which agrees with the topological criterion.
In comparison, when $\nu_{x}$ ($\nu_{y}$) is trivialized by the exchange field,
the two Majorana zero modes at the dislocations with $\textbf{B}=\pm\bm{e_x}$ ($\pm\bm{e_y}$) disappear,
while the two Majorana zero modes at the dislocations with $\textbf{B}=\pm\bm{e_y}$ ($\pm\bm{e_x}$)
remain intact; this is clearly shown in the inset on top of the red (blue) region in Fig.\ref{fig5}. We can draw an important conclusion from Fig.\ref{fig5}. Namely, when the chemical potential is close to $\mu_{c,-}$, tuning the direction of a weak exchange field
can control the presence or absence of Majorana zero modes at a dislocation. These results
suggest that the unique spin-splitting band structure induced by AM and Rashba SOC provides a basis for the realization of controllable Majorana modes.

\section{Discussions and conclusions}\label{IV}

The combination of $d$-wave AM and Rashba SOC
leads to a unique spin-splitting band structure respecting the $C_{4z}\mathcal{T}$ symmetry.
Breaking the $C_{4z}\mathcal{T}$ symmetry by an exchange field, we find that both
the strong topology and the weak topology of the resulting band structure
show a sensitive dependence on the exchange field's
direction. As a consequence of the sensitive dependence of weak topological indices
on the exchange field's direction, we find that the presence or absence of topological bound states
at a dislocation can be easily controlled by adjusting the exchange field's direction.
By putting the 2D altermagnetic metal in proximity to an $s$-wave superconductor, we find a
similar sensitive dependence of the band topology and dislocation bound states
on the exchange field when the chemical potential is appropriately chosen.
As the dislocation bound states are Majorana zero modes in this case, their sensitivity towards external fields implies high degrees of controllability; this may make the superconducting altermagnetic materials stand out as platforms to detect and manipulate Majorana zero modes.

From a symmetry perspective, the change  from one set of topological dislocation bound states  into their $C_{4z}$-rotational counterparts upon reversing the exchange field's direction
is a direct manifestation of the $C_{4z}\mathcal{T}$ symmetry of the pristine altermagnetic system.
In the light of this, an observation of this switching behavior of topological dislocation bound states can be taken as a strong signature
of $d$-wave AM. In experiments, the field-sensitive dislocation bound states are
robust due to their topological origin and can be easily detected by scanning tunneling microscopy. {\zb Up to now, 
a list of 2D materials with $d$-wave AM have been predicted\cite{Bai2024review}, which makes the experimental 
realization of our proposal quite promising. Particularly, we note that $d$-wave AM has recently predicted
to be a possible origin of the time-reversal symmetry breaking observed in FeSe grown on SrTiO$_{3}$\cite{Mazin2023monolayer}.
Our findings in the current paper may provide a way to test this prediction in this intensively-studied material.} To conclude, 
given the ubiquitous presense of topological defects in real materials and a continuously accumulated list of candidates for altermagnetic materials,
our findings unveil a promising route to an effective experimental diagnosis of AM in these materials.\\

\section*{Acknowledgements}

D. Z., D. L., Z.-Y. Z, and Z. Y. are supported by the National Natural Science Foundation of China (Grant No. 12174455), Natural Science Foundation of Guangdong Province
(Grant No. 2021B1515020026), and Guangdong
Basic and Applied Basic Research Foundation (Grant No.
2023B1515040023). Z. W. is supported by National Key R$\&$D Program of China (Grant No. 2022YFA1404103), NSFC (Grant No.~11974161) and Shenzhen Science and Technology Program (Grant No.~KQTD20200820113010023).

\bibliography{dirac}

\end{document}